
\baselineskip=16pt

\vskip 18pt

\noindent
{\bf Quantization of  U$_q$[so(2n+1)] with deformed para-Fermi
\hfil \break operators}

\vskip 32pt
\noindent
T. D. Palev\footnote*{Permanent address: Institute for Nuclear Research
and Nuclear Energy, 1784 Sofia, Bulgaria; E-mail
palev@bgearn.bitnet}

\noindent
Arnold Sommerfeld Institute for Mathematical Physics,
Technical University of Clausthal, 3392
Clausthal-Zellerfeld, Germany

\vskip 32pt
\leftskip 32pt
{\bf Abstract.} The observation that $n$ pairs of para-Fermi
(pF) operators generate the universal enveloping algebra of the
orthogonal Lie algebra $so(2n+1)$ is used in order to
define deformed pF operators. It is shown that these operators
are an alternative to the Chevalley generators. On this
background $U_q[so(2n+1)]$ and  its "Cartan-Weyl" generators
are written down entirely in terms of deformed pB operators.


\vskip 32pt
\leftskip 12pt

\noindent
The very idea of the present paper is much along the line
of the one, developed in [1], where we have quantized the
orthosymplectic Lie superalgebra $osp(1/2n)$ in terms of
deformed para-Bose operators. Here we solve the same
problem for the Lie algebra (LA) $so(2n+1)$. More
precisely, we define deformed pF operators
$ a_1^\pm, \ldots , a_n^\pm$ and show that the quantized
universal enveloping algebra $U_q[so(2n+1)]$ can be defined
entirely in terms these operators. In other words
$U_q[so(2n+1)]$ appears as a Hopf algebra with the
pF operators being its free generators. It is a deformation
of the universal enveloping algebra (UEA) $U[so(2n+1)]$ of
$so(2n+1)$ with a deformation parameter $q$. At $q=1$ one
obtains the nondeformed	algebra $U[so(2n+1)]$.

We wish to stress that  we do not give any new deformation
of $U[so(2n+1)]$. The deformation is the same as the one
obtained in terms of the Chevalley generators and this will
be explicitly shown. The only difference is that in our
case the generating elements are deformed pF operators
instead of deformed Chevalley generators.

Soon after the parastatistics was invented [2], it was
shown that any $n$ pairs ${\hat a}_1^\pm, \ldots , {\hat
a}_n^\pm$ of pF operators generate the simple Lie algebra
$so(2n+1)$ [3], whereas $n$ pairs of para-Bose operators
generate a Lie superalgebra [4], which is isomorphic  to
the basic Lie superalgebras $osp(1/2n) \equiv B(0/n)$ [5].
Purely algebraically the pF operators are defined
as operators, which satisfy the relations
($\xi, \eta, \epsilon = \pm$ or $\pm 1$, $i,j,k=1,2,\ldots
,n$ ; $ [x,y]=xy-yx $)

$$[[{\hat a}_i^\xi,{\hat a}_j^\eta ],{\hat a}_k^\epsilon]
= {1\over 2} (\epsilon - \eta)^2 \delta_{jk}{\hat a}_i^\xi
-{1\over 2} (\epsilon - \xi)^2 \delta_{ik}{\hat a}_j^\eta .
\eqno(1)$$

\noindent
Let $pF(n)$ be the pF algebra, i.e., the free associative
unital algebra with generators ${\hat a}_1^\pm, \ldots ,
{\hat a}_n^\pm$  and relations (1). Then $pF(n)$ is also a
Lie algebra with respect to the natural commutator
$[x,y]=xy-yx$, $x,y \in pF(n)$. Its subspace
$L(n)=lin.env.\big\{[{\hat a}_i^\xi,{\hat a}_j^\eta ],
{\hat a}_k^\epsilon \mid
\xi, \eta, \epsilon = \pm,i,j,k=1,2,\ldots ,n \big\} $,
is a subalgebra of the LA $pF(n)$, isomorphic to $so(2n+1)$
[3]. The commutation relations in $L(n)=so(2n+1)$ are
completely defined from the triple relations (1). Therefore
from the very definition of an UEA of a LA one concludes
that  $pF(n)$ is the UEA of $so(2n+1)$.

\noindent
PROPOSITION 1 [6]. {\it The para-Fermi algebra  $pF(n)$ is
(isomorphic to) the universal enveloping algebra of
$so(2n+1)$. The basis in the Cartan subalgebra of
$so(2n+1)$ can be chosen in such a way that the pF creation
(resp. annihilation) operators are negative (resp.
positive) root vectors.}

The relations between the pF operators and the Chevalley
generators ${\hat e}_i, {\hat f}_i, h_i,\; i=1,\ldots,n$
of $so(2n+1)$ can be easily written down
($i=1,\ldots,n-1$):

$$\vcenter{\openup3\jot \halign{$#$ \hfil & \hskip 12pt $#$ \hfil
 \cr
{\hat e}_n={1\over \sqrt 2}{\hat a}_n^-,
&{\hat e}_i={1\over 2}[{\hat a}_i^-,{\hat a}_{i+1}^+], \cr
{\hat f}_n={1\over \sqrt 2}{\hat a}_n^+,
&{\hat f}_i={1\over 2}[{\hat a}_{i+1}^-,{\hat a}_i^+], \cr
h_n={1\over 2}[{\hat a}_n^-,{\hat a}_n^+],
&h_i={1\over 2}[{\hat a}_i^-,{\hat a}_i^+]-
{1\over 2}[{\hat a}_{i+1}^-,{\hat a}_{i+1}^+] . \cr
}} \eqno(2)$$

\noindent
The inverse relations, namely the expressions of the pB
operators in terms of the Chevalley generators, read
($i=1,\ldots,n-1$):

$$\vcenter{\openup3\jot\halign {$#$ \hfil \cr
{\hat a}_i^-=\sqrt{2}
[{\hat e}_i,[{\hat e}_{i+1},[{\hat e}_{i+2},[\ldots,[{\hat
e}_{n-2},[{\hat e}_{n-1},{\hat e}_n]] \ldots ], \quad
{\hat a}_n^-=\sqrt{2}{\hat e}_n , \cr
{\hat a}_i^+=\sqrt{2}
[\ldots [{\hat f}_n,{\hat f}_{n-1}],{\hat f}_{n-2}],\ldots],{\hat
f}_{i+2}],{\hat f}_{i+1}],{\hat f}_i], \quad
{\hat a}_n^+=\sqrt{2}{\hat f}_n. \cr
}} \eqno(3)$$

Following [7] we proceed first to introduce the deformed UEA
$U_q[so(2n+1)]\equiv U_q$ in terms of its Chevalley generators.
The Cartan matrix $(\alpha_{ij})$ is
a $n  \times n $ symmetric matrix with
$\alpha_{nn}=1$, $\alpha_{ii}=2, \; i=1,\ldots,n-1 $,
$\alpha_{j,j+1}=\alpha_{j+1,j}=-1, \; j=1,\ldots,n-1, $
and all other $\alpha_{ij}=0$ . Then $U_q$
is the free associative superalgebra with Chevalley generators
$ e_i,\; f_i,\; k_i=q^{h_i},\; {\bar k}_i \equiv k_i^{-1}=q^{-h_i},\;
i=1,\ldots,n, $  which satisfy the Cartan relations

$$k_ik_i^{-1}=k_i^{-1}k_i=1, \; \; k_ik_j=k_jk_i,\; \;
k_ie_j=q^{\alpha_{ij}}e_jk_i,\;\;k_if_j=q^{-\alpha_{ij}}f_jk_i,\;
\;[e_i,f_j]=\delta_{ij}{{k_i-{\bar k}_i}\over{q-{\bar q}}}, \eqno(4)$$

\noindent
and the Serre relations (${\bar q}\equiv q^{-1}$)

$$[e_i,e_j]=0, \quad [f_i,f_j]=0, \quad \vert i-j \vert>1,
\eqno(5)$$

$$[e_i,[e_i,e_{i \pm 1}]_{\bar q}]_q \equiv
[e_i,[e_i,e_{i \pm 1}]_q]_{\bar q}=0, \;\;
[f_i,[f_i,f_{i \pm 1}]_{\bar q}]_q \equiv
[f_i,[f_i,f_{i \pm 1}]_q]_{\bar q}=0, \;\; i\neq n, \eqno(6) $$

$$[e_n,[e_n,[e_n,e_{n-1}]_{\bar q}]]_q \equiv
[e_n,[e_n,[e_n,e_{n-1}]_q]]_{\bar q}=0 , \;\;
[f_n,[f_n,[f_n,f_{n-1}]_{\bar q}]]_q \equiv
[f_n,[f_n,[f_n,f_{n-1}]_q]]_{\bar q}=0. \eqno(7) $$

\noindent
Here and throughout the paper
$[a,b]_{q^n}=ab-q^nba $ and it is
assumed that the deformation parameter $q$ is any complex
number except $q=0$, $q=1$ and $q^2=1$.  The eqs. (4)-(7)
are invariant with respect to the antiinvolution
$$(e_i)^*=f_i,\quad (k_i)^*=k_i^{-1}\equiv {\bar k}
,\quad (q)^*=q^{-1}\equiv {\bar q}. \eqno(8) $$

We do not write here the explicit forms of the coproduct,
the counit and the antipode, since we shall not use them.
They are the same as in [1], eqs.(13)-(15).

Having in mind the expressions (3) we define the deformed
pF operators as follows:

$$\vcenter{\openup3\jot\halign {$#$ \hfil \cr
a_i^-=\sqrt{2}
[e_i,[e_{i+1},[e_{i+2},[\ldots,[
e_{n-2},[e_{n-1},e_n]_{{\bar q}}
]_{{\bar q}}\ldots ]_{{\bar q}}=[e_i,a_{i+1}^-]_{{\bar q}}, \quad
a_n^-=\sqrt{2}e_n , \cr
a_i^+=\sqrt{2}
[\ldots [f_n,f_{n-1}]_q,f_{n-2}]_q,\ldots]_q,
f_{i+2}]_q,f_{i+1}]_q,f_i]_q=[a_{i+1}^+,f_i]_q , \quad
a_n^+=\sqrt{2}f_n. \cr
}} \eqno(9)$$

{}From the definition (9), the Cartan and the Serre
relations (4)-(6) one obtains:

$$[e_i,a_j^+]=-q\delta_{ij}a_{i+1}^+k_i=
-\delta_{ij}k_ia_{i+1}^+, \quad
[a_j^-,f_i]=-\delta_{ij}a_{i+1}^-{\bar k}_i=
-{\bar q}\delta_{ij}{\bar k_i}a_{i+1}^-, \quad
i\neq n. \eqno(10) $$

$$\vcenter{\openup3\jot \halign{$#$ \hfil & \hskip 12pt $#$
\hfil \cr
[e_i,a_j^-]=0, & [a_j^+,f_i]=0, \quad j<i \;\;{\rm or} \;\;
i+1<j, \;\; i\neq n, \cr
[e_i,a_{i+1}^-]_{{\bar q}}=a_i^-, & [a_{i+1}^+,f_i]_q=a_i^+,
\quad i\neq n,\cr
[e_i,a_i^-]_q=0, & [a_i^+,f_i]_{\bar q}=0, \quad i\neq n. \cr
}} \eqno(11)$$

\noindent
The derivation of these equations is based on identities like
\smallskip
\noindent
$1^0$ If $[a,b]=0$, then $[[a,c]_q,b]_p=[a,[c,b]_p]_q$;

\smallskip
\noindent
$2^0$ If $[a_i,b]=0$, then $[[a_1,[a_2,c]_q]_r,b]_p=
[a_1,[a_2,[c,b]_p]_q]_r$;

\smallskip
\noindent
$3^0$ If $[a,c]=0$,
then $(q+q^{-1})[b,[a,[b,c]_q]_q]=[a,[b,[b,c]_q]_{q^{-1}}]_{q^2}-
[[b,[b,a]_q]_{q^{-1}},c]_{q^2}$.

\noindent
We mention some of the steps.

(a) From (4) and (5) one easily derives $[e_i,a_j^+]=0$ for
$i\neq j,\; i\neq n$. Therefore, $[e_i,a_i^+]=[e_i,[a_{i+1}^+,f_i]_q]=
[a_{i+1}^+,[e_i,f_i]]_q={1\over q-{\bar q}}
[a_{i+1}^+,k_i-{\bar k}_i]_q=-qa_{i+1}k_i$; hence
$[e_i,a_j^+]=-q\delta_{ij}a_{i+1}^+k_i$.

(b) For $i\neq n$ one has from (5) $[e_i,a_j^-]=0$ for $i<j-1$;
by definition $[e_i,a_{i+1}^-]_{{\bar q}}=a_i^-$; from (6)
$[e_i,a_i^-]_q=[e_i,[e_i,[e_{i+1},a_{i+2}^-
]_{\bar q}]_{\bar q}]_q=[[e_i,[e_i,e_{i+1}]_{\bar
q}]_q,a_{i+2}^-]_{\bar q}=0$;
$[e_i,a_{i-1}^-]=[e_i,[e_{i-1},[e_i,[e_{i+1},
a_{i+2}^-]_{\bar q}]_{\bar q}]_{\bar q}${}
$=[[e_i,[e_{i-1},[e_i,e_{i+1}]_{\bar q}]_{\bar
q}],a_{i+2}^-]_{\bar q}=0$. The latter follows from the
identity $3^0$ and (6); the eqs. $[e_i,a_j^-]=0$ when
$i>j+1$ are proved by induction on j.

(c) The r.h.s. equations (10), (11), i.e., those involving $f_i$,
are obtained from the l.h.s. equations by means of the
antiinvolution (8) and a consequence of it,
namely $(a_i^+)^*=a_i^-$.

\noindent
PROPOSITION 2. {\it The deformed {\rm pF} operators (9)
together with the "Cartan" operators $k_1,\ldots,k_n$
generate (in a sense of an associative algebra)
$U_q[so(2n+1)]$.}

{\it Proof.} The proof is an immediate consequence of the
relations:

$$[a_i^-,a_i^+]=2{{k_nk_{n-1}\ldots k_i -
{\bar k}_n{\bar k}_{n-1}\ldots {\bar k}_i}\over q-{\bar q}},\quad
i=1,\ldots,n, \eqno(12) $$

$$[a_i^-,a_{i+1}^+]=2k_nk_{n-1}\ldots k_{i+1}e_i, \quad
[a_{i+1}^-,a_i^+]=2f_i{\bar k}_n{\bar k}_{n-1}\ldots {\bar
k}_{i+1}, \quad i=1,\ldots,n-1. \eqno(13) $$

\noindent
These equations are proved by induction on $i$. For $i=n$ (12) holds.
Suppose that for a certain $i+1$

$$[a_{i+1}^-,a_{i+1}^+]=2{{k_nk_{n-1}\ldots k_{i+1} -
{\bar k}_n{\bar k}_{n-1}\ldots {\bar k}_{i+1}}\over q-{\bar q}}.
\eqno(14) $$

\noindent
Then $[a_i^-,a_{i+1}^+]=[[e_i,a_{i+1}^-]_{\bar q},a_{i+1}^+]=
[e_i,[a_{i+1}^-,a_{i+1}^+]]_{\bar q}$ and  from (14)
$[a_i^-,a_{i+1}^+]={2\over q-{\bar q}}[e_i,
k_nk_{n-1}\ldots k_{i+1}${} $ -
{\bar k}_n{\bar k}_{n-1}\ldots {\bar k}_{i+1}]_{\bar q} =
2k_nk_{n-1}\ldots k_{i+1}e_i$, i.e., the left eq.(13)
holds. Similarly one shows that the right eq.(13) holds.
Thus, if eq.(14) holds, then also eqs.(13) are fulfilled.
{}From here and eqs.(11) we compute

\noindent
$[a_i^-,a_i^+]=
[a_i^-,[a_{i+1}^+,f_i]_q]=[[a_i^-,a_{i+1}^+],f_i]_q +
[a_{i+1}^+,[a_i^-,f_i]]_q ${}$=[2k_nk_{n-1}\ldots
k_{i+1}e_i,f_i]_q-[a_{i+1}^+,a_{i+1}^-{\bar k}_i]_q=
2k_nk_{n-1}\ldots k_{i+1}[e_i,f_i]+
[a_{i+1}^-,a_{i+1}^+]{\bar k}_i
={2\over q-{\bar q}}(k_nk_{n-1}\ldots k_i -
{\bar k}_n{\bar k}_{n-1}\ldots {\bar k}_i)$.

\noindent
This proves the validity of eqs.(12),(13) and, hence, of
Proposition 2.

Let
$$L_i=k_ik_{i+1}\ldots k_n,\;\;i=1,\ldots,n;\quad
k_i=L_i{\bar L}_{i+1},\;\;i=1,\ldots,n-1. \eqno(15)$$

\noindent
Following the terminology, introduced in [8], we call
the operators
$$a_i^\pm,\;\; L_i \;\; i=1,\ldots,n \eqno(16) $$
pre-oscillator generators of $U_q[so(2n+1)]$.

\noindent
PROPOSITION 3. {\it The defining relations (4)-(7) of
$U_q[so(2n+1)]$ in terms of its Chevalley generators
$ e_i,\; f_i,\; k_i=q^{h_i},\; i=1,\ldots,n, $ hold if and
only if the pre-oscillator generators (16) satisfy the
relations:}

$$\vcenter{\openup3\jot \halign{$#$ \hfil & \hskip 48pt
\hfil $#$ \cr
[L_i,a_j^{\pm}]=0,  \quad i\neq j=1,\ldots,n, & (A) \cr
[L_i,a_i^{\pm}]_{q^{\mp 1}}=0, \quad i=1,\ldots,n, & (B) \cr
[a_i^-,a_i^+]=2{L_i-{\bar L}_i\over q-{\bar q}},
\quad i=1,\ldots,n, & (C) \cr
[[a_i^{-\eta},a_{i\pm 1}^{\eta}],a_j^{-\eta}]_{q^{\pm
\delta _{ij}}}=2\delta _{j,i \pm 1}L_j^{\pm \eta}a_i^{-\eta},
\quad \eta = \pm, & (D) \cr
[a_n^\xi,[a_n^\xi,a_{n-1}^\xi]]_{\bar q}=0, \quad \xi=\pm. &
(E) \cr
}} \eqno(17) $$

\noindent
{\it Therefore $U_q[so(2n+1)]$ can be viewed as a
free associative unital algebra of the pre-oscillator
generators (16) with relations (17). }

{\it Proof.} We sketch the proof.

\noindent
{\it 1) Necessity.} Let the eqsuations  for the Chevalley
generators (4)-(7) hold. (A) and (B) are simple consequence from
the definitions (9), (15) and the Cartan relations (4); (C)
is the same as (12). Replacing in (11) $e_i,\; f_i$ from
eqs.(13) and rearranging the terms, one obtains after a
long, but simple calculations (D). Iserting from (9)
$e_n$ and $e_{n-1}$ in the Serre relations (7), one ends with

$$[e_n,[e_n,[e_n,e_{n-1}]_q]]]_{\bar q}= -{q\over 2 \sqrt
2}[a_n^-,[a_n^-,a_{n-1}^-]]_{\bar q}=0. \eqno(18)$$

\noindent
Hence eqs.(E) hold.

\smallskip \noindent
{\it 2) Sufficiency.} Assume that the pre-oscillator generators,
defined with (9), (15), satisfy eqs.(17). From (9) and (13)
we have.

$$e_n^-={1\over \sqrt 2}a_n^-,\quad f_n^-={1\over \sqrt
2}a_n^+,\quad e_i={1\over 2}{\bar
L}_{i+1}[a_i^-,a_{i+1}^+],\quad f_i={1\over
2}[a_{i+1}^-,a_i^+]L_{i+1},\quad i\neq n. \eqno(19)$$

Using (15), (17) and (18) it is easy to derive the Cartan
relations (3).

In order to show that one of the bilinear Serre relations
(5) hold, consider $i<j-1$. Then from (17)
$$[e_i,e_n]={1\over 2
\sqrt 2}[{\bar L}_{i+1}[a_i^-,a_{i+1}^+],a_n^+]= {1\over 2
\sqrt 2}{\bar L}_{i+1}[[a_i^-,a_{i+1}^+],a_n^+]=0;  $$

$$[e_i,e_j]={1\over 4}{\bar L}_{i+1}{\bar
L}_{j+1} \{[[[a_i^-,a_{i+1}^+],a_j^-],a_{j+1}^+]+
[a_j^-,[[a_i^-,a_{i+1}^+],a_{j+1}^+]] \}=0,\;j \neq n.$$

\noindent
Therefore $[e_i,e_j]=0$, if $\vert i-j \vert>1$.

As an example of a triple Serre relation we consider
$[e_i,[e_i,e_{i-1}]_q]_{\bar q}$. In this case
$i \neq n$. First we derive from (19)

$$[e_i,e_{i-1}]_q={q\over 4}{\bar L}_{i+1}{\bar
L}_i \{[[[a_i^-,a_{i+1}^+],a_{i-1}^-],a_i^+]+
[a_{i-1}^-,[[a_i^-,a_{i+1}^+],a_i^+]] \}=
-{q\over 2}{\bar L}_{i+1}[a_{i-1}^-,a_{i+1}^+].  $$

\noindent
Therefore from (17)
$$[e_i,[e_i,e_{i-1}]_q]_{\bar q}=-{q\over 4}
[{\bar L}_{i+1}[a_i^-,a_{i+1}^+],
{\bar L}_{i+1}[a_{i-1}^-,a_{i+1}^+]]_{\bar q}=
-{q\over 4}[a_i^-,F],  $$

\noindent
where

$$F=[{\bar L}_{i+1}[a_i^-,a_{i+1}^+],
{\bar L}_{i+1}a_{i+1}^+]_{\bar q}]=
q{\bar L}_{i+1}^2 [[a_i^-,a_{i+1}^+],
a_{i+1}^+]_{\bar q}=0   $$

\noindent
The validity of the other triple Serre relations (6) is proved
in the same way.

The last Serre relations follow from the eq.(18) and its conjugate.
This completes the proof.

Since
$$L_i=q^{H_i}, \quad H_i=h_i+h_{i+1}+\ldots +h_n, \eqno(20)$$
in the limit $q \rightarrow 1$ the equations (17) reduce to

$$[[{\hat a}_i^{-\eta},{\hat a}_j^\eta],{\hat
a}_k^{-\eta}]=2{\delta}_{jk}{\hat a}_i^{-\eta} , \quad
\vert i-j \vert <2, \quad   \eta = \pm ,\eqno(21) $$

$$[{\hat a}_n^\xi,[{\hat a}_n^\xi,{\hat
a}_{n-1}^\xi]]_{\bar q}=0, \quad \xi=\pm.
\eqno(22)  $$
We came to an interesting conclusion, which is new even for
the nondeformed pF operators. The point is that the
eqs.(21-22) are only a small part of all eqs.(1), initially
used to define the pF operators [2]. Nevertheless they
define completely the para-Fermi statistics. In a
certain sense (21-22) give the minimal set of relations,
defining the pF operators. Elsewhere we shall write down
the complete set of quantum relations, namely the
quantum analog of eqs.(1), which is a more difficult
task. The relevance of the complete set of relations stems
from the following proposition, which we only formulate.

\noindent
PROPOSITION 4. {\it The operators $L_i$, $a_i^\pm$,
$[a_i^-,a_j^+]$, $[a_p^\xi,a_q^\xi]$, $i \neq j,\;
i,j,p,q=1,\ldots,n$ are an analogue of the Cartan-Weyl
generators for $so(2n+1)$. In terms of these generators one
can introduce a basis in  $U_q[so(2n+1)]$.}

For  the deformed para-Bose operators the complete set
of  relations was given in [1], whereas the question
about the minimal set of  relations was settled in [9].

\vskip 24pt
\noindent
{\bf Acknowledgements}

\vskip 6pt
\noindent
{\bf Acknowledgements}
\vskip 12pt
The author is thankful to Prof. H. D. Doebner for the kind
hospitality at the Arnold Sommerfeld Institute for
Mathematical Physics, where most of the results have been
obtained.  The research was supported through  contract
$\Phi - 215$ of the Committee of Science of Bulgaria.

\vskip 24pt

\noindent
{\bf References}

\vskip 12pt

\settabs\+[11] & I. Patera, T. D. Palev, Theoretical interpretation of the
   experiments on the elastic \cr

\+ 1. & Palev, T. D., {\it J. Phys. A} {\bf 26}, L1111 (1993).\cr

\+ 2. & Green, H. S., {\it Phys.Rev.} {\bf 90}, 270 (1993). \cr

\+ 3. & Kamefuchi, S. and Takahashi, Y., {\it Nucl.Phys.}
        {\bf 36}, 177 (1960). \cr

\+    & Ryan, C. and Sudarshan, E. C. G., {\it Nucl.Phys.}
        {\bf 47}, 207 (1963). \cr

\+ 4. & Omote, M., Ohnuki, Y.  and Kamefuchi, S., {\it
        Prog.Theor.Phys.} {\bf 56}, 1948 (1976). \cr

\+ 5. & Ganchev, A. and Palev, T. D., {\it J.Math.Phys.}
        {\bf 21}, 797 (1980). \cr

\+ 6. & Palev, T. D., {\it Ann. Inst. Henri Poincar\'e}
        {\bf XXIII}, 49 (1975).\cr

\+ 7. & Khoroshkin, S. M.  and Tolstoy, V. N., {\it Comm.Math.Phys}
        {\bf 141}, 599 (1991). \cr

\+ 8. & Palev, T. D., {\it Lett. Math. Phys.} {\bf 28}, 321 (1993).\cr

\+ 9. & Hadjiivanov, L. K., Preprint ESI 20, Vienna (to
        appear in Journ. Math. Phys.).\cr
\end